\documentclass[aps,prb,twocolumn,superscriptaddress,amsmath,amssymb]{revtex4}

\usepackage{graphicx}
\usepackage{dcolumn}
\usepackage{booktabs}
\usepackage{bm}
\usepackage{color}
\usepackage{ulem}

\newcommand{\Tc}{\ensuremath{T^*}}
\newcommand{\GMT}{Ge$_{1-x}$Mn$_x$Te}
\newcommand{\xa}{\ensuremath{x_{\rm a}}}
\newcommand{\xICP}{\ensuremath{x_{\rm ICP}}}
\newcommand{\MTK}{\ensuremath{M_{\rm 7T,2K}}}

\begin{document}

\title{Enhanced ferromagnetic transition temperature induced by a microscopic structural rearrangement in the diluted magnetic semiconductor Ge$_{\bm{1-x}}$Mn$_{\bm{x}}$Te}

\author{M.~Kriener}
\email[corresponding author: ]{markus.kriener@riken.jp}
\author{T.~Nakajima}
\author{Y.~Kaneko}
\author{A.~Kikkawa}
\author{D.~Hashizume}
\affiliation{RIKEN Center for Emergent Matter Science (CEMS), Wako 351-0198, Japan}
\author{K.~Kato}
\affiliation{RIKEN SPring-8 Center, Hyogo 679-5148, Japan}
\author{M.~Takata}
\affiliation{RIKEN SPring-8 Center, Hyogo 679-5148, Japan}
\author{T.~Arima}
\affiliation{RIKEN Center for Emergent Matter Science (CEMS), Wako 351-0198, Japan}
\affiliation{Department of Advanced Materials Science, University of Tokyo, Kashiwa 277-8561, Japan}
\author{Y.~Tokura}
\affiliation{RIKEN Center for Emergent Matter Science (CEMS), Wako 351-0198, Japan}
\affiliation{Department of Applied Physics and Quantum-Phase Electronics Center (QPEC), University of Tokyo, Tokyo 113-8656, Japan}
\author{Y.~Taguchi}
\affiliation{RIKEN Center for Emergent Matter Science (CEMS), Wako 351-0198, Japan}

\date{\today}

\begin{abstract}
The correlation between magnetic properties and microscopic structural aspects in the diluted magnetic semiconductor \GMT\ is investigated by x-ray diffraction and magnetization as a function of the Mn concentration $x$. The occurrence of high ferromagnetic-transition temperatures in the rhombohedrally distorted phase of slowly-cooled \GMT\ is shown to be directly correlated with the formation and coexistence of strongly-distorted Mn-poor and weakly-distorted Mn-rich regions. It is demonstrated that the weakly-distorted phase fraction is responsible for the occurrence of high-transition temperatures in \GMT. When the Mn concentration becomes larger, the Mn-rich regions start to switch into the undistorted cubic structure, and the transition temperature is suppressed concurrently. By identifying suitable annealing conditions, we successfully increased the transition temperature to above 200~K for Mn concentrations close to the cubic phase. Structural data indicate that the weakly-distorted phase fraction can be restored at the expense of the cubic regions upon the enhancement of the transition temperature, clearly establishing the direct link between high-transition temperatures and the weakly-distorted Mn-rich phase fraction.
\end{abstract}

\maketitle

\section{Introduction}
GeTe is a very intriguing material attracting huge interest for its diversity on different physical properties: It is a fairly good conductor due to native Ge vacancies creating hole-type charge carriers, otherwise a narrow-gap semiconductor.\cite{edwards06a} It features a many-valley band structure\cite{herman68a,ciucivara06a,edwards06a} and was among the first semiconductors with such a peculiar band structure found to superconduct after the prediction by Cohen back in the 1960s.\cite{cohen64a,hein64a,finegold64a} At high temperatures, it crystallizes in a cubic structure (space group $Fm\bar{3}m$; $\beta-$GeTe). Upon cooling, the system undergoes a structural transition into a rhombohedral phase (space group $R3m$; $\alpha-$GeTe) at approximately 430$^\circ$C due to a polar distortion which leads to an elongation of the unit cell along the cubic $[111]_{\rm c}$ (or rhombohedral $[003]_{\rm h}$ in hexagonal setting) direction.\cite{goldak66a,pawley66a,steigmeier70a,chattopadhyay87a,rabe87b,schlieper99a,fons10a,kolobov14a,dyang16a} In recent years, GeTe attracted attention due to the prediction of a giant Rashba-spin splitting in its bulk bands due to the pronounced polar structure,\cite{disante13a,picozzi14a} for which shortly after experimental evidence was reported.\cite{rinaldi14a,krempasky15a,liebmann16a} 

Doped GeTe compounds also serve as base materials in thermoelectricity\cite{levin13a,jklee14a} as, e.g., (Ge,Pb,Yb)Te,\cite{jqli15a} GeTe:Bi$_2$Te$_3$,\cite{dwu14a} or the well-known family of TAGS (GeTe)$_{x}$(AgSbTe$_2$)$_{1-x}$ compounds.\cite{snyder08a,davidow13a} Another important feature from the viewpoint of application is the functionality of GeTe-based alloys in switchable phase-change-memory devices.\cite{jongenelis96a,nonaka00a,kolobov04a,kolobov14a,upadhyay14a} GeTe is one end member of the GeTe\,--\,Sb$_2$Te$_3$ pseudo-binary system where state-of-the-art phase-change materials are found.\cite{lencer08a} The switching between crystalline and amorphous phases can be induced by, e.g., laser irradiation (optical pulse) or electrical fields (electric pulse) causing an order\,--\,disorder transition in analogy with liquid-crystal transitions, where a supercooled disordered state, i.e., a glassy amorphous state, competes with a long-range crystallographically ordered state.\cite{chen86a,kolobov04a}

GeTe also offers the possibility of enhanced magnetic interactions and applicability in spintronics devices: Magnetism is induced in GeTe when doping with Cr, Mn, or Fe at the Ge site,\cite{cochrane73a,cochrane74a,fukuma03b,tong11a} adding the possibility of multiferroicity to its list of features.\cite{przybylinska14a,kriegner16a} These doped materials belong to the family of binary diluted magnetic semiconductors, such as (Ga,Mn)N or (Ga,Mn)As, for which magnetic ordering temperatures around room temperature or above were theoretically predicted to occur at Mn concentrations of 5\% or 10\%, respectively, but have not yet been experimentally realized.\cite{dietl00a,jungwirth05a,dietl14a,lchen11a} Among these materials, \GMT\ attracted much interest since Mn easily replaces Ge, forming single-phase GeTe\,--\,MnTe alloys up to $x\gtrsim 0.5$.\cite{lechner10a} Upon Mn doping, the polar distortion reduces and eventually the system is driven back to its cubic structure. Upon further doping, the crystal structure gradually changes to hexagonal (space group $P6_{3}/mmc$), and the end compound MnTe is an antiferromagnet. 

In an early work, Cochrane \textit{et al.} reported a linearly increasing ferromagnetic-transition temperature $T_{\rm c}(x)$ with a maximum value of about 165~K for $x = 0.5$,\cite{cochrane74a} and discussed the emergence of carrier-mediated ferromagnetism in bulk \GMT\ in a Rudermann-Kittel-Kasuya-Yoshida (RKKY) framework. Since then much work has been done on thin films of \GMT, with $T_{\rm c}$ being enhanced due to an increase of the charge-carrier concentration.\cite{cochrane74a,fukuma03a,hsato05a,chen08a,fukuma08a,lechner10a,hassan11a} The highest values of $T_{\rm c}$ so-far reported in this system are around $\sim 190- 200$~K.\cite{fukuma08a,hassan11a} However, the doping concentrations $x$, for which such high-$T_{\rm c}$ values were reported, vary widely in the available literature, and are often in contradiction to the early work by Cochrane \textit{et al.},\cite{cochrane74a} although epitaxial strain will modify $T_{\rm c}(x)$ in thin films to some extent.

In a preceding publication (Ref.~[\onlinecite{kriener16a}]), we investigated this system, and revealed that there are in fact two distinct magnetic phases in the low-doped region ($\xa\lesssim 0.15$; for the definition of \xa\ see below) of the phase diagram, depending on the heat treatment of the samples. These phases are characterized by very different onset temperatures \Tc\ of magnetization (as measured in $B=0.1$~T), explaining qualitatively the contradicting results found in literature. Moreover we demonstrated that a sample can be switched back and forth from its low-\Tc\ to the high-\Tc\ phase by performing the different heat-treatment procedures alternatingly. Hence \GMT\ is the magnetic analog of the aforementioned structural phase-change materials. 

The high-\Tc\ phase is formed when a sample is cooled slowly and in a controlled way from about 900~K, which is approximately in the middle between melting point and structural phase transition, i.e., where the system is already solidified but still in its high-temperature cubic phase.\cite{StruktComment} The phase diagram exhibits a dome-like structure with maximum values of \Tc\ of $\sim 180$~K around $\xa=0.075$ (see Fig.~\ref{fig1}). By contrast, when a sample is quenched from the high-temperature cubic phase, substantially smaller values of \Tc\ are found: \Tc\ is reduced by a factor of five to six around the maximum of the dome. Also, the \Tc\ of quenched samples is roughly proportional to $x$ as it is expected for RKKY-like ferromagnetic order and in agreement with the earlier work (Ref.~[\onlinecite{cochrane74a}]). Around $\xa\approx 0.15 - 0.22$, the system gradually changes into its cubic structure at room temperature, and above this doping-induced structural-transition range, there is no difference in \Tc\ found any more between samples grown by either cooling recipe. In this doping range, the slope of the \Tc\ vs \xa\ curve is also linear but reduced as compared to the rhombohedral low-\Tc\ phase. 

We presented evidence that a different degree of Mn inhomogeneity caused by a spinodal decomposition during the cooling process is at work in this system and responsible for the rhombohedral high-\Tc\ phase. In the latter the spinodal decomposition leads to the formation of \GMT\ with a spatial-position-dependent Mn concentration, i.e., Mn-rich regions embedded in a matrix of low-doped or even pristine GeTe, while in the low-\Tc\ phase the Mn dopants are much more homogeneously distributed. The characteristic size of these regions was estimated to be a few tens of nm.\cite{kriener16a} Here we present a detailed structural analysis based on high-resolution synchrotron x-ray diffraction (SXRD) data. In our previous study, the complicated structural situation in the high-\Tc\ phase was only qualitatively and briefly discussed based on the roughly estimated lattice constants for one controlled-cooled sample. The present results are based on detailed Rietveld refinements of various heat-treated samples and provide quantitative evidence that in the high-\Tc\ phase the degree of the rhombohedral distortion changes with $x$ in the Mn-rich regions and that these are responsible for the emergence of high-\Tc-values. Upon further doping, cubic phase fractions develop locally in the Mn-rich regions, leading to a reduction of \Tc. Eventually the structure changes globally to cubic, and the difference in \Tc\ depending on the heat treatment disappears. We also report newly identified heat-treatment conditions to enhance \Tc, which was successfully increased to $\sim 214$~K for Mn concentrations in the structural-transition range from rhombohedral to cubic at room temperature. 

The paper is organized as follows. In Sec.~\ref{Exp}, experimental procedures are described. In Sec.~\ref{PhaDi}, we present an extended phase diagram including the newly established ``maximum-\Tc'' phase line. The subsequent Sec.~\ref{optimized} focusses on optimizing the heat treatment of samples to further enhance \Tc. The results of the high-resolution SXRD experiments on selected samples are summarized in Sec.~\ref{XRDpatterns}. In Sec.~\ref{discussion}, we discuss the role of the Mn redistribution and the rhombohedrally distorted structure, based on the results of Rietveld refinements of the SXRD data and its impact on \Tc. Sec.~\ref{summary} summarizes the paper. Additional discussion and supplemental data are provided in Ref.~[\onlinecite{Suppl}].

\section{Experimental}\label{Exp}
All the \GMT\ samples investigated in this paper are bulk polycrystals which were grown by conventional melt-growth and Bridgman methods. Stoichiometric amounts of GeTe (purity: 5N) and MnTe (3N+) were mixed, sealed into evacuated quartz tubes, and subsequently heated to about $1073-1223$~K. They were kept at this temperature for at least 12~h. Then they were slowly cooled through the melting point (GeTe: $T_{\rm melt}\approx 1000$~K, Ge$_{0.5}$Mn$_{0.5}$Te: $T_{\rm melt}\approx 1073$~K)\cite{johnston61a} to about 900~K where the batches are solidified in the undistorted cubic high-temperature GeTe structure. Then the batches were either quenched or slowly cooled down to room temperature ($-5$~K/h). These two heat treatments yield the aforementioned different magnetic phases. As for batches grown by Bridgman method, the upper heater was set to $1073-1223$~K, the lower to $623-723$~K. The sealed quartz tubes with the mixed powder were again kept for at least 12~h at the higher temperature and subsequently slowly lowered (2~mm/h) and either slowly cooled down to room temperature or quenched when the quartz tube was at approximately 900~K. Additional annealing attempts at 900~K before either quenching or slow cooling did not have any effect on the magnetic phases. Magnetic sample characterization was carried out with commercial magnetometers (MPMS XL and MPMS-3, Quantum Design). The temperature-dependent XRD patterns were taken with a commercial in-house diffractometer (Rigaku SmartLab) in N$_2$ atmosphere to prevent oxidization of the powder samples. The high-resolution synchrotron radiation experiments were performed at the BL44B2 beam line at SPring-8.\cite{kato16a} The Rietveld refinements were performed using the software RIETAN-FP.\cite{izumi07a} Inductively coupled plasma atomic-emission spectroscopy-based (ICP-AES) analyses of the chemical composition of selected samples were done at Hitachi Power Solutions Co., Ltd.

\section{Results}

\subsection{Extended phase diagram}
\label{PhaDi}
\begin{figure}[t]
\centering
\includegraphics[width=8.5cm,clip]{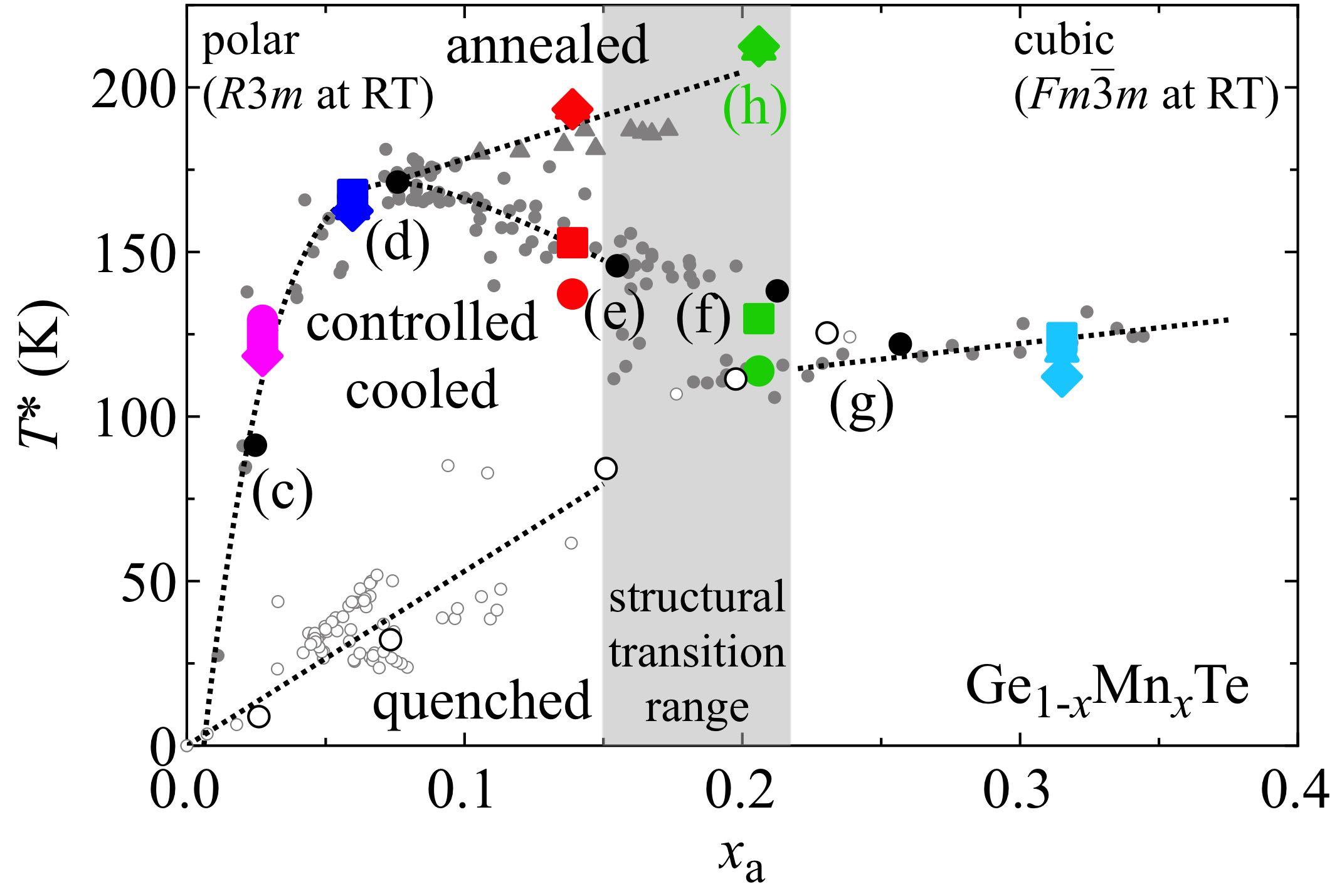}
\caption{Extended phase diagram of \GMT\ including data of annealed samples which exhibit higher \Tc\ values than observed before. Filled symbols refer to the \Tc\ of controlled-cooled (filled circles) or annealed (filled triangles), open symbols to quenched samples, respectively. Colored symbols of various shapes identify samples for which different heat treatments were applied, see text. Pairs of quenched and controlled-cooled samples with similar Mn concentrations for which high-resolution SXRD measurements were carried out are indicated with black open or filled circle symbols, respectively. The labels (c) to (g) next to them refer to panels in Fig.~\ref{fig4} where the respective data are shown. In addition, SXRD data on the maximum-\Tc\ sample, labeled (h) and indicated by a filled green diamond, are also shown in Fig.~\ref{fig4}. Some of the data points were reproduced from Ref.~[\onlinecite{kriener16a}]. The dotted lines are guides to the eyes and the gray-shaded area indicates the structural-transition range. The latter is only an approximation and does not indicate exact phase boundaries (see text and Sec.~S9 in the Supplemental Material Ref.~\onlinecite{Suppl}} for details).
\label{fig1}
\end{figure}

Figure~\ref{fig1} presents an extended phase diagram of \Tc\ plotted against \xa\ with \Tc\ being defined as the intersection of the tangent to the linear part of $M(T)$ data with the temperature axis as measured in $B=0.1$~T under field-cooling condition. For an example, see the red-dotted line for one data set in Fig.~\ref{fig3}~(a). Here we use the notation \Tc\ instead of $T_{\rm c}$ because this temperature was determined from data measured in an applied field which smears out the transition and leads to an overestimation of the thermodynamic $T_{\rm c}$. For two test samples (a controlled-cooled and a quenched rhombohedral one), we determined the thermodynamic $T_{\rm c}$ via Arrot-Noakes plots and also measured the temperature-dependent magnetization upon warming in zero field after field cooling in $B=0.1$~T, cf.\ Fig.~S1 in the Supplemental Material [\onlinecite{Suppl}]. The phase diagram in Fig.~\ref{fig1} is modified from the phase diagram presented in our previous publication [\onlinecite{kriener16a}]. There, the transition temperature of a sample was plotted against its magnetic moment \MTK\ measured at $T=2$~K and $B=7$~T, and the corresponding Mn concentrations were labeled $x_m$. These were calculated from \MTK\ under the assumption that all Mn$^{2+}$ ions contribute with their full moment $5\,\mu_{\rm B}$, and hence $x_m$ was taken as a measure of the effective Mn concentration. This approach was chosen because as-grown batches suffer from a slight gradient of the Mn concentration on a cm scale, leading to deviations from the nominal Mn concentration for a certain sample cut from these batches. 
Since then we chemically analyzed selected samples throughout the phase diagram by inductively coupled plasma atomic-emission spectroscopy (ICP-AES) to determine Mn concentrations \xICP. For small $x_m$, there is little difference between $x_m$ and \xICP. However, upon increasing $x_m$ the difference grows. Therefore the ICP-AES results were used to adjust $x_m$ of all samples to obtain $\xa=\xICP(x_m)$ as plotted in Fig.~\ref{fig1}. The experimental results of the ICP-AES experiments and the details of the adjustment procedure are summarized in the Supplemental Material [\onlinecite{Suppl}], Fig.~S2. We use these \xa\ throughout this paper.\cite{xacomment} Compared to our preceding publication, the main change here is the slight rescaling of the horizontal axis in Fig.~\ref{fig1}. In addition, the position of the gray-shaded structural-transition range is modified. Its location is based on new results obtained from the structural analysis presented in this paper (see Sec.~\ref{discussion}). 

In Fig.~\ref{fig1}, filled symbols refer to controlled-cooled (filled circles) or annealed (filled triangles), open symbols to quenched samples. Some of the data points plotted with small gray circles are reproduced from Ref.~[\onlinecite{kriener16a}]. In the rhombohedral part of the phase diagram, the samples exhibit a huge difference in \Tc, depending on the heat treatment. In the structural-transition range, the cubic phase stabilizes more and more, and for larger Mn concentrations samples realize the same \Tc\ independent of the heat treatment for a given Mn concentration. Various data points in Fig.~\ref{fig1} are highlighted by thicker black and colored symbols. The thicker filled and open circle symbols in black indicate samples which were examined by high-resolution SXRD experiments, and they are labeled pairwise (c), ..., (g). Each pair consists of one controlled-cooled and one quenched sample with similar \xa. The labels refer to panels in Fig.~\ref{fig4} where relevant SXRD data on these pairs of samples are shown (see Sec.~\ref{XRDpatterns}). The colored symbols of various shapes indicate test samples which were annealed under certain conditions to possibly enhance \Tc. These annealing experiments will be discussed next. For the sample with the highest \Tc\ found in this study [labeled (h)], the SXRD data are also shown in Fig.~\ref{fig4}. The main result here is the emergence of higher-\Tc\ values exceeding 200~K, establishing the new ``maximum-\Tc'' line labeled as ``annealed'' in Fig.~\ref{fig1}.

\subsection{Optimizing the heat-treatment procedure}
\label{optimized}
\begin{figure}[t]
\centering
\includegraphics[width=7.5cm,clip]{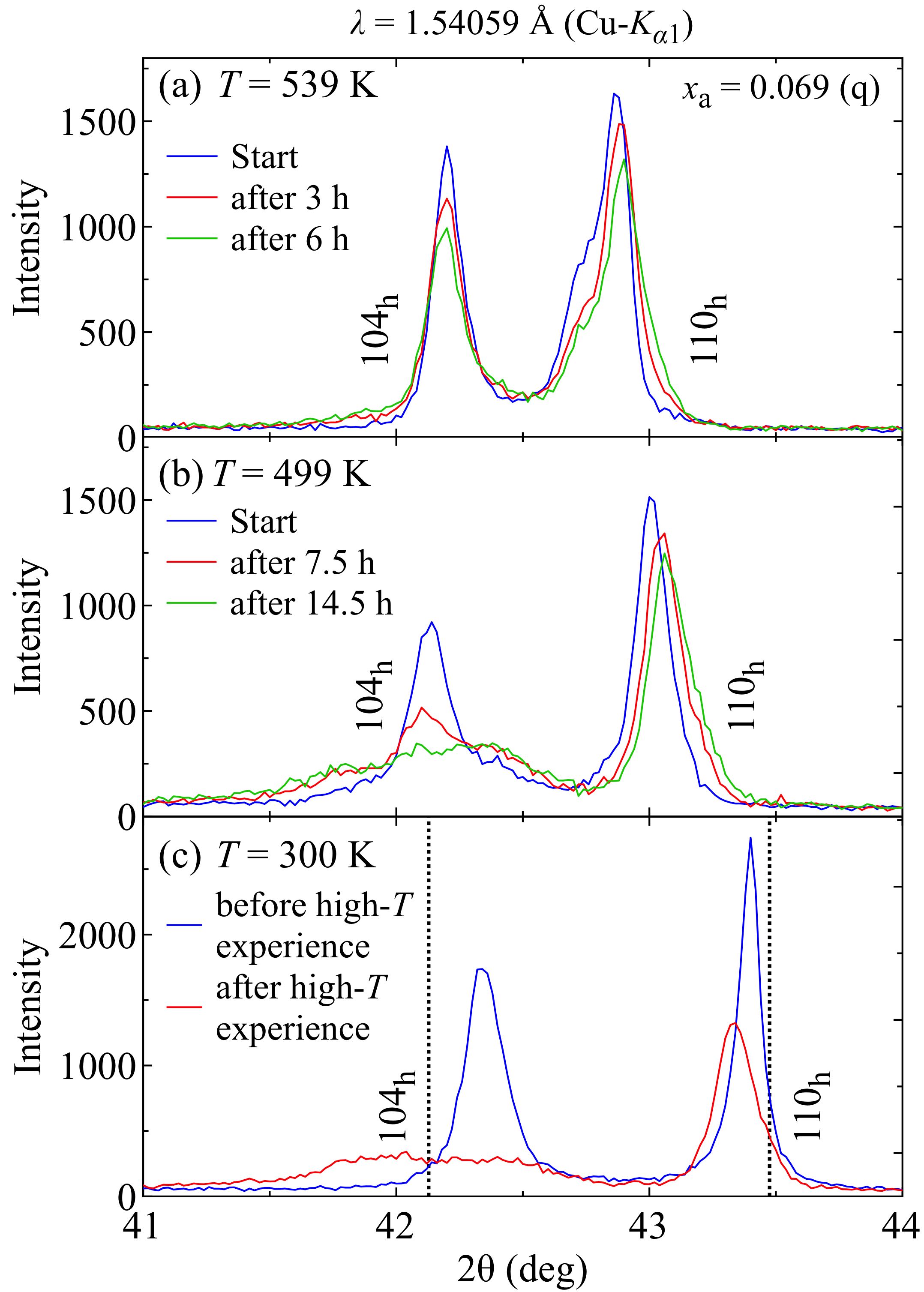} 
\caption{Temperature-dependent XRD patterns around the $104_{\rm h}$ and $110_{\rm h}$ reflections (in hexagonal setting) of an initially quenched sample with $\xa=0.069$. First, the powder sample was heated up to about 900~K into its high-temperature cubic phase. Then the temperature was slowly lowered. (a) XRD patterns taken at $T=539$~K as a function of elapsed time after the temperature had settled. (b) XRD patterns taken as a function of elapsed time, after lowering the temperature to $T=499$~K. (c) Comparison of XRD data taken at room temperature before and after the high-temperature experience. The dashed lines indicate the position of both reflections for pristine GeTe. The different positions $2\theta$ of the XRD peaks in the three panels are due to the temperature-induced change of the lattice constants (see text for details).} 
\label{fig2}
\end{figure}
Figure~\ref{fig2} summarizes temperature-dependent XRD data taken with an in-house diffractometer. The employed radiation is Cu-$K_{\alpha 1}$ with a wavelength of $\lambda = 1.54059$~\AA. The plots show an expanded view on the $2\theta$ range around the $104_{\rm h}$ and $110_{\rm h}$ reflections (in hexagonal setting). The $104_{\rm h}$ peak is sensitive to changes of the degree of the rhombohedral distortion, i.e., a sharp $104_{\rm h}$ peak with strong intensity implies a homogeneous distortion throughout the pulverized powder sample. By contrast, a broad peak and / or even multiple peaks indicate that different parts of the sample exhibit rhombohedral distortions of different degree. The $110_{\rm h}$ peak reflects the $d$ spacing perpendicular to the direction of the polar distortion.\cite{UCellcomment} The data shown in Fig.~\ref{fig2} were measured on a quenched sample with $\xa=0.069$. Before heating up the sample, XRD data was taken at room temperature [blue data in Fig.~\ref{fig2}~(c)]. Subsequently, the sample temperature was increased to about 900~K. At selected temperatures the XRD pattern was measured during the warming run. Some of the scans are shown in Fig.~S3 of the Supplemental Material Ref.~[\onlinecite{Suppl}]. The temperature-induced first-order structural phase transition upon warming took place between 650~K and 670~K. At 900~K the XRD pattern was measured roughly every hour, in total three times.\cite{XRDtimecomment1} There is only very little change in the intensity of the diffraction profile (see Ref.~[\onlinecite{Suppl}]). Next, the sample temperature was decreased and XRD data was taken at certain temperatures. 

The crystal structure started to distort again between 620~K and about 580~K. At $T=539$~K, the XRD pattern was taken seven times roughly every hour. The initial (blue data), the fourth (red), and the last measurement (green) taken at 539~K are shown in Fig.~\ref{fig2}~(a). There is no strong effect on the width of the diffraction lines and only a small reduction of the intensity of the peaks at this temperature. The most apparent change here is the weakening and disappearance of the small shoulder visible at the low-angle side of the $110_{\rm h}$ peak by time. The shoulder might be due to small remaining cubic-phase fractions in the pulverized powder sample, possibly indicating that the structural phase transition was not fully completed yet when stabilizing at this temperature upon cooling. While waiting at this temperature, the remaining cubic-phase fraction of the sample also switched to the low-temperature rhombohedral phase and hence the shoulder disappeared. 

Next, the temperature was lowered to 499~K and the XRD pattern was measured in total 16 times in $\sim 60 - 80$~minutes intervals.\cite{XRDtimecomment2} Three selected XRD patterns are shown in Fig.~\ref{fig2}~(b). At this temperature, a strong effect on the shape and intensity of the $104_{\rm h}$ peak is observed. After $\sim 14.5$~h, the $104_{\rm h}$ peak had shrunk and broadened significantly, indicating that the Mn ions had redistributed and, depending on the local Mn concentration on the $\sim 10$~nm scale, locally different distortion angles had developed throughout the sample. It is noted that the overall rhombohedral symmetry of the structure is kept intact and that only the distribution of the Mn ions, and hence the degree of rhombohedral distortion, has changed. Thus we conclude that $\sim 500$~K is the optimum temperature for the spinodal decomposition to proceed in this sample. 

Eventually the sample temperature was lowered back to room temperature and the XRD pattern was measured once again. The latter data are plotted along with data taken at room temperature before the high-temperature experience in Fig.~\ref{fig2}~(c) for comparison. The difference due to the spinodal decomposition is striking. The two vertical dotted lines at $2\theta \approx 42.1^{\circ}$ and $43.5^{\circ}$ mark the positions of the two reflections for pristine GeTe. Apparently the intensity of the $104_{\rm h}$ reflection shifted partially even \textit{below} the corresponding peak position for $\xa = 0$. This indicates that some Mn-poor regions of the sample are even more strongly rhombohedrally distorted than pure GeTe. In addition, the $110_{\rm h}$ peak was also affected by the heat treatment. It became much weaker and somewhat broader compared to the initial pattern, indicating that the $d$ spacing perpendicular to the polar axis also got affected.  
\begin{table}[t]
\centering
\begin{ruledtabular}
\caption{Time-dependent annealing effect at $T=500$~K on as-grown controlled-cooled samples of \GMT. Their \Tc\ values are indicated by colored symbols in the phase diagram in Fig.~\ref{fig1}. The symbols used therein for the different annealing steps are shown in brackets in the first line. The colors used are with increasing \xa: magenta, blue, red, green, cyan. The temperatures \Tc\ are given in K, \MTK\ in $\mu_{\rm B}/{\rm f.u.}$}
\label{tab1}
\begin{tabular}{ccccccccc}
\toprule
\multicolumn{3}{c}{as grown} ($\bullet$)   & \multicolumn{2}{c}{1 day} ($\scriptstyle\blacksquare$) &  \multicolumn{2}{c}{1 week} ($\blacktriangle$)  &   \multicolumn{2}{c}{3 weeks} ($\scriptstyle\blacklozenge$)\\ \addlinespace[0.1em]
 \xa\  & \Tc\ & \MTK\ & \Tc\ & \MTK\ & \Tc\ & \MTK\ & \Tc\ & \MTK  \\ \hline \addlinespace[0.75em]
 0.027 & 129  & 0.090 & 125  & 0.091 & 121  & 0.091 & 118  & 0.090 \\ \addlinespace[0.1em] 
 0.060 & 167  & 0.189 & 168  & 0.191 & 163  & 0.190 & 162  & 0.190 \\ \addlinespace[0.1em] 
 0.139 & 137  & 0.400 & 153  & 0.413 & 194  & 0.394 & 193  & 0.395 \\ \addlinespace[0.1em] 
 0.206 & 114  & 0.557 & 130  & 0.581 & 212  & 0.467 & 214  & 0.491 \\ \addlinespace[0.1em] 
 0.315 & 121  & 0.782 & 124  & 0.846 & 120  & 0.787 & 112  & 0.712 \\ \addlinespace[0.1em] 
\bottomrule
\end{tabular}
\end{ruledtabular}
\end{table}
\begin{figure}[t]
\centering
\includegraphics[width=8.5cm,clip]{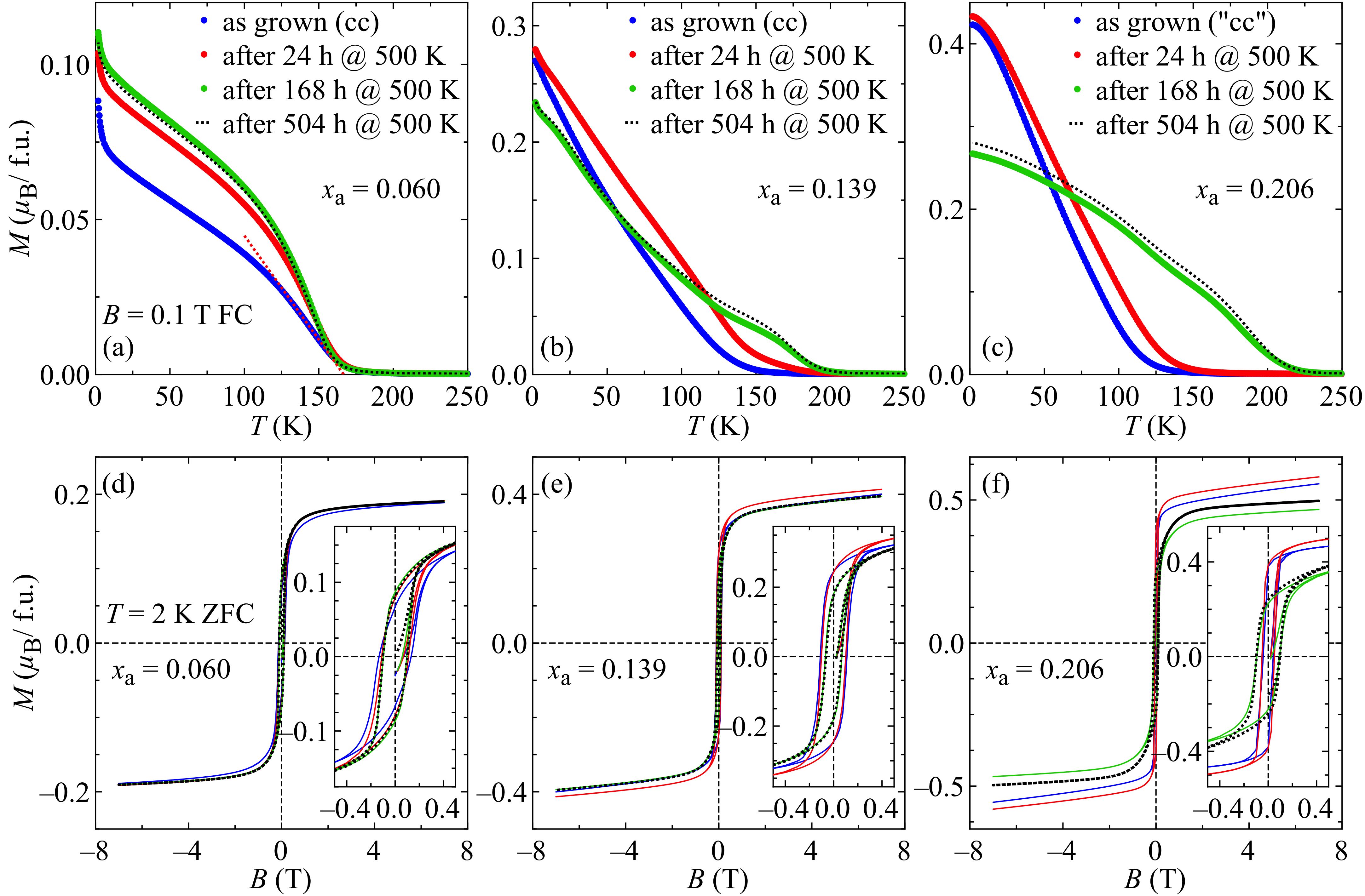}
\caption{(a) -- (c) Magnetization in $B=0.1$~T measured after annealing experiments on three different samples with (a) $\xa=0.060$, (b) $\xa=0.139$, and (c) $\xa=0.206$. Temperature-dependent data (field cooled; FC) is shown for the as-grown controlled-cooled samples (blue data points) and after annealing for one day (red), one week (green), and three weeks (black dotted lines) at the optimal spinodal decomposition temperature of 500~K. The lower panels (d)\,--\,(f) summarize the field dependences of the magnetization at $T=2$~K after each annealing step of the samples shown in the upper panels (a)\,--\,(c) (zero-field cooled; ZFC), see text. The red-dotted line in panel (a) indicates exemplarily the definition of \Tc.}
\label{fig3}
\end{figure}

Having identified the optimal annealing temperature at which the spinodal decomposition takes place most efficiently, the next questions which arise are: (a) What is the ideal annealing time, i.e., how long does it take for the rearrangement of the Mn ions? and (b) Does such a heat treatment further enhance \Tc? To address these questions, four controlled-cooled as-grown test samples from the rhombohedral phase region and one from the cubic one were selected. Three subsequent annealing experiments were carried out on these samples: First each sample was sealed into an evacuated quartz tube and heated up to 893~K ($=620^{\circ}$C, i.e., into the cubic phase). Then the samples were kept at 893~K for 2~h, cooled down slowly to 500~K, and kept at that temperature for one day. After cooling down to room temperature, the magnetization of the samples was measured. Afterwards, they were sealed again, heated up to 893~K, slowly cooled to 500~K, kept there for one week, and measured. In the last annealing experiment, the samples were kept for three weeks at 500~K.\cite{Tanealcomment1} Sometimes annealing slightly affected a sample's surface color, but the bulk was not degraded, as also discussed in Sec.~\ref{anneff}. Table~\ref{tab1} summarizes the \Tc\ and \MTK\ values of these samples as grown and after each annealing experiment. Their positions in the phase diagram are indicated by colored symbols in Fig.~\ref{fig1} (with increasing \xa: magenta, blue, red, green, cyan). The \Tc\ values for the controlled-cooled as-grown samples are plotted with circle symbols. The corresponding data points referring to one day, one week, and three weeks of annealing are marked with square, triangle, and diamond symbols, respectively, as also indicated in brackets in the top line of Table~\ref{tab1}.

The magnetization data for three out of the five test samples are summarized in Fig.~\ref{fig3}. Data for the most lightly- and heavily-doped samples are not shown here for simplicity. These can be found in the Supplemental Material Ref.~[\onlinecite{Suppl}], Fig.~S4. Figures~3(a)\,--\,3(c) contain the temperature dependence [$M(T)$], and Figs.~3(d)\,--\,3(f) the field dependence [$M(B)$] of the magnetization, respectively.  The annealing process affects the samples in different ways: We observe changes in \Tc\ in $M(T)$ measurements as well as changes in the shape of the hysteresis loops and of the magnetic moment \MTK\ in $M(B)$ measurements. We note that the $M(B)$ curves are not fully saturated yet at 7~T and in higher-doped samples their slope can slightly vary upon annealing. The phase diagram can be divided into three regions and in the following, we will discuss these regions separately.

Light doping $\xa\lesssim 0.075$: Below the peak concentration of the dome of the high-\Tc\ phase, annealing does not have a strong impact on \Tc. For both examined samples with $\xa = 0.027$ and 0.060, \Tc\ remains roughly constant and tends to decrease slightly in longer annealing experiments ($\xa = 0.027$: $-8.5$\%, $\xa = 0.060$: $-3.0$\% after three weeks annealing). The effect on the hysteresis loops is also tiny. They become somewhat sharper around the origin while the magnetic moment \MTK\ at 7~T and 2~K remains unchanged.

When approaching the structural-transition range ($0.075 \lesssim\ \xa\lesssim 0.220$; gray-shaded area in Fig.~\ref{fig1}), a clear annealing effect on \Tc\ is observed. The test sample with $\xa=0.139$ was found to exhibit a higher \Tc\ value by about 40\% after annealing for one week. The strongest enhancement of \Tc\ was observed for the sample with $\xa=0.206$, where \Tc\ reached 212~K after one week annealing, corresponding to an increase of 86\%. For both samples, the even longer three-weeks-annealing experiment did not change \Tc\ significantly any more (193~K and 214~K, respectively), which suggests that about one week annealing at 500~K is sufficient to reach the fully spinodal-decomposed state in the sense of maximizing \Tc. We note that it is possible to obtain any \Tc\ value between the ``as grown'' controlled-cooled and the ``annealed'' phase line in Fig.~\ref{fig1} as already suggested by the scatter of the data points in the phase diagram. The gray triangle data points in Fig.~\ref{fig1} indicate additional samples which were annealed at 500~K, but for some of them the respective \Tc\ values are below the dotted line denoted as ``annealed'', which is in this sense the ``maximum-\Tc'' line observed in this study.\cite{Tanealcomment2} As for $M(B)$ measurements, annealing causes again a slight sharpening of the hysteresis loops. However, in contrast to the lower-doped samples, here the magnetic moment at 7~T and 2~K is not constant any more. Upon annealing an increase as well as a decrease was found to be possible. 

High doping $\xa\gtrsim 0.220$: The \Tc\ of the test sample with $\xa=0.315$ in the cubic phase is almost unaffected by the different annealing experiments (see Fig.~S4~(b) in Ref.~[\onlinecite{Suppl}]) in agreement with the phase diagram. The hysteresis loops of this sample exhibit only tiny modifications, but the observed changes in \MTK\ after each successive annealing experiment are the largest among all test samples. Apparently, these changes are not correlated with the annealing time, but they increase systematically with the Mn concentration [$\Delta \MTK=0.018$ ($\xa = 0.139$), $0.114$ ($\xa = 0.206$), and $0.134$ ($\xa =0.315$)], cf.\ Table~\ref{tab1}. We will address this issue in the Discussion Sec.~\ref{discussion} in more detail.

\subsection{Evolution of XRD patterns} 
\label{XRDpatterns}
\begin{figure}[t]
\centering
\includegraphics[width=8.5cm,clip]{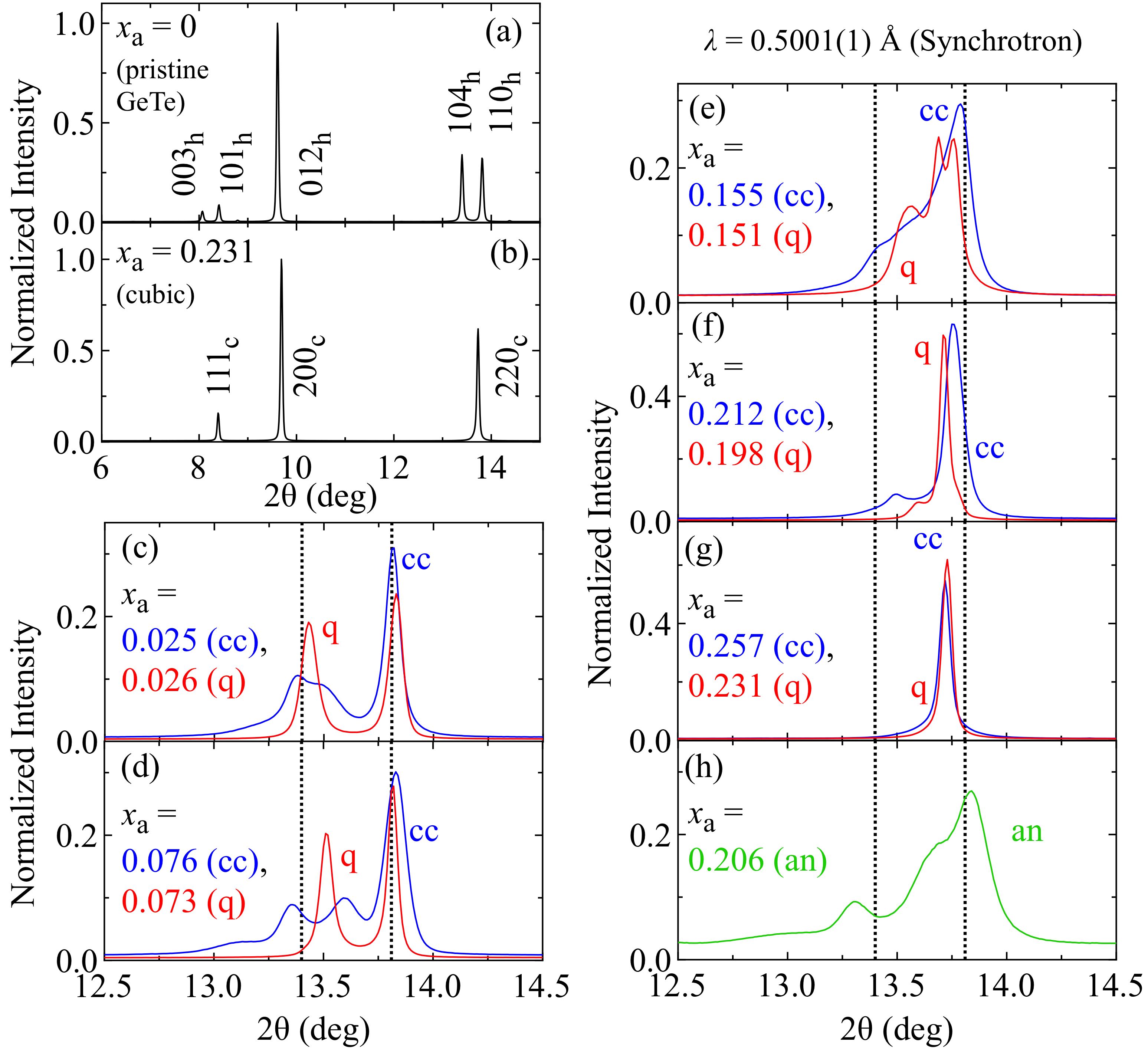} 
\caption{High-resolution SXRD patterns at room temperature. (a) Typical peak profile for pristine rhombohedral GeTe (space group $R3m$). (b) Corresponding data for a cubic \GMT\ sample with $\xa=0.231$ (space group $Fm\bar{3}m$). In the cubic phase all peaks shown in the panel are single peaks. (c)\,--\,(h) SXRD data around the rhombohedral $104_{\rm h}$ and $110_{\rm h}$ reflections for selected samples. The corresponding data points in the phase diagram in Fig.~\ref{fig1} are indicated by the same labels (c)\,--\,(h). Each panel (c)\,--\,(g) contains data for a controlled-cooled (labeled ``cc'', blue) and a quenched (``q'', red) sample with similar Mn concentrations for comparison. Panel (h) shows the data for the annealed (``an'', green) sample with the highest \Tc, see text for details. The vertical dotted lines in the panels indicate the positions of the $104_{\rm h}$ and $110_{\rm h}$ reflections in pristine GeTe.}
\label{fig4}
\end{figure}

Figure~\ref{fig4} summarizes SXRD measurements at room temperature for selected samples from different parts of the phase diagram. The wavelength used in this experiment is $\lambda = 0.5001(1)$~\AA, i.e., different from that of the Cu-$K_{\alpha 1} $ radiation used in the in-house experiments, and therefore the $2\theta$ values in Fig.~\ref{fig4} differ from those presented in Fig.~\ref{fig2}. In total, we took high-resolution SXRD data on nine quenched, eight controlled-cooled, and three annealed samples of \GMT. In addition, pristine (rhombohedral) GeTe was also measured. The data of the latter are shown in Fig.~\ref{fig4}~(a) for comparison. The corresponding pattern of the cubic material is shown in Fig.~4(b) measured on a sample from the heavily-doped part of the phase diagram ($\xa = 0.231$). The main reflection $200_{\rm c}$ (or $012_{\rm h}$ in the rhombohedral phase) is almost unchanged in either structure, while each of the peaks on both sides of the main reflection are split in the distorted rhombohedral structure. 

The data of the samples which are highlighted with thicker black symbols in Fig.~\ref{fig1} are shown in Figs.~4(c) to 4(g) with increasing \xa. To compare how the different heat treatments affect the patterns, each of these panels contains data for a quenched (``q''; red) and a similarly-doped controlled-cooled sample (``cc''; blue). Among them, parts of the data presented in (d) (``cc'' data) and (g) (``q'' data) were already shown in Ref.~[\onlinecite{kriener16a}]. Figure~4(h) presents data taken on the annealed sample (``an''; green) which exhibits the highest \Tc\ found in this study (green diamond symbol in Fig.~\ref{fig1}). As in Fig.~\ref{fig2}, we focus on the rhombohedral $104_{\rm h}$ and $110_{\rm h}$ peaks. The vertical dotted lines in these panels mark the positions of the $104_{\rm h}$ and $110_{\rm h}$ peaks for pristine GeTe for comparison ($2\theta \approx 13.4^{\circ}$ and $13.8^{\circ}$, respectively). We will describe the changes with \xa\ step by step while moving from Figs.~4(c) to 4(h). 

In Fig.~\ref{fig4}~(c), the SXRD data of the quenched sample ($\xa=0.026$; red) exhibits two well-separated $104_{\rm h}$ and $110_{\rm h}$ peaks indicating a rather uniform rhombohedral distortion throughout the pulverized powder sample. By contrast, the $104_{\rm h}$ reflection of the controlled-cooled sample ($\xa=0.025$; blue) is broadened, showing a splitting into at least two peaks and a pronounced smaller peak intensity. The latter is apparently distributed over a larger angular range, i.e., the rhombohedral distortion is \textit{not} uniform in this sample. There are rather various regions with different degrees of distortion while the overall rhombohedral symmetry is preserved. This observation is in accord with the inhomogeneous distribution of the Mn atoms with a characteristic length scale of several tens of nm, as presented in Fig.~4 of our preceding publication Ref. [\onlinecite{kriener16a}], which may also contribute to the observed broadening of the diffraction peaks.

Figure~\ref{fig4}~(d) shows the SXRD data for samples with Mn concentrations $\xa = 0.073$ (``q'') and 0.076 (``cc''), i.e., they are located in the center of the dome of controlled-cooled samples in the phase diagram in Fig.~\ref{fig1}. Again there are two comparably sharp and well separated peaks visible in the SXRD data of the quenched sample (red). The lower $104_{\rm h}$ peak has shifted to higher angles in agreement with the larger Mn concentration. In the respective data for the controlled-cooled sample (blue), the $104_{\rm h}$ peak has split as indicated by the two maxima. Moreover there is a broad hump (or third maximum) at the low-angle side. The lower maximum and the broad hump are clearly below the vertical dotted line at $2\theta\approx 13.4^{\circ}$, indicating that some parts of the sample are even more strongly distorted than pure GeTe. On the other hand, the upper maximum of the $104_{\rm h}$ peak has shifted towards higher angles than the corresponding single peak in the data of the quenched sample, in agreement with the aforementioned scenario towards the formation of Mn-rich less-distorted regions (closer to cubic) in a surrounding matrix of almost pristine GeTe. The $110_{\rm h}$ peak is also somewhat broader. 

Figure~\ref{fig4}~(e) contains data for samples which are located around the lower end of the gray-shaded structural-transition range in Fig.~\ref{fig1}: $\xa = 0.151$ (``q''; red) and $\xa = 0.155$ (``cc''; blue). In this Mn concentration range, the $104_{\rm h}$ peak of the quenched sample has also broadened and started to merge with the $110_{\rm h}$ peak, indicating the formation of cubic phase fractions. The $110_{\rm h}$ peak has split into two maxima, most likely due to a mixture of the cubic $220_{\rm c}$ and the rhombohedral $110_{\rm h}$ reflections. In the case of the controlled-cooled sample, the features at the low-angle side below the dotted line have almost vanished. Most intensity has shifted towards the $104_{\rm h}$ peak as indicated by its broad left shoulder.

Figure~\ref{fig4}~(f) contains data for samples with $\xa = 0.198$ (``q''; red) and $\xa = 0.212$ (``cc''; blue). Both samples are located in the upper half of the gray-shaded structural-transition range in Fig.~\ref{fig1}. In case of the quenched sample, the lower $104_{\rm h}$ peak has almost vanished but is still discernible. The main part of the sample is already cubic, as indicated by the strong $110_{\rm h}$ (or $220_{\rm c}$) peak. The situation for the controlled-cooled sample is similar. The intensity of the lower $104_{\rm h}$ peak became weak and it has developed into a single feature with only little intensity below the dotted line at about $13.4^{\circ}$, while the $110_{\rm h}$ ($220_{\rm c}$) reflection is strong. However, both peaks are still more separated than in the case of the quenched sample. 

Figure~\ref{fig4}~(g) summarizes data for a quenched ($\xa=0.231$; red) and a controlled-cooled ($\xa=0.257$; blue) sample which are located above the gray-shaded structural-transition range in Fig.~\ref{fig1}. In both cases, there is only a single sharp cubic $220_{\rm c}$ peak, and no shoulders, humps, or similar features are visible. Both samples have completed the doping-induced structural phase transition. In the cubic phase, there is no apparent spinodal decomposition taking place, and hence the microscopic crystal structure is the same and independent of the heat treatment, in agreement with the results of magnetic measurements.\cite{kriener16a} 

Figure~\ref{fig4}~(h) contains data for the long-time annealed sample with the highest \Tc\ observed here. The Mn concentration is  $\xa =0.206$ in the gray-shaded structural-transition range in Fig.~\ref{fig1}. The $104_{\rm h}$ reflection exhibits doubly split maxima, the lower one below the dotted line at $2\theta \approx 13.4^{\circ}$ and the upper one has partially merged with the $110_{\rm h}$ reflection. There is also a broad feature at lower angles below the lower maximum of the $104_{\rm h}$ reflection. As for the Mn concentration, this panel locates between panels (e) and (f). In the data of the controlled-cooled samples shown in the latter two panels, the intensity of the $104_{\rm h}$ reflection has mostly shifted above the lower dotted line. By contrast, the data in panel (h) rather resembles qualitatively the data of the controlled-cooled sample shown in panel (d), which has a much smaller Mn concentration \xa\ and a $104_{\rm h}$ reflection that has split into two maxima and a broad feature below these. In spite of the large \xa, the annealing procedure has switched this sample back into a structural condition which was found in as-grown controlled-cooled samples located around $\xa \approx 0.075$, i.e., in the center of the dome of the high-\Tc\ phase. The apparent difference, beside the larger \xa, is that the annealed sample has an enhanced \Tc. This will be discussed in more detail in the light of lattice constants and the evolution of the spinodal decomposition in the next section. 

\section{Discussion} 
\label{discussion}
We carried out Rietveld refinements for all samples measured by synchrotron radiation and estimated their lattice constants $a_{\rm h}$ and $c_{\rm h}$ (in hexagonal setting) for the rhombohedral phase and $a_{\rm c}$ for the cubic phase. The results are summarized in Fig.~\ref{fig5}~(a). For a better comparison, the hexagonal parameters are plotted in their pseudocubic setting: $\tilde{a}_c =\sqrt{2}a_{\rm h}$ and $\tilde{c}_c = c_{\rm h}/\sqrt{3}$. The lattice parameters of quenched samples are indicated with open symbols, those of controlled-cooled and annealed samples with filled symbols ($\tilde{a}_c$: circles, $\tilde{c}_c$: squares, $a_{\rm c}$: triangles; for the color code see below). Figure~\ref{fig5}~(b) summarizes the corresponding volume fractions of the respective phases. The Supplemental Material provides raw SXRD and Rietveld-refined data for selected samples, see Fig.~S5 in Ref.~[\onlinecite{Suppl}]. As in Fig.~\ref{fig1}, the gray-shaded area in Fig.~\ref{fig5}~(a) indicates the structural-transition range where the system changes to cubic structure.

\subsection{Rietveld refinement and lattice constants}
\subsubsection{Quenched rhombohedral and cubic samples}
\begin{figure}[t]
\centering
\includegraphics[width=8cm,clip]{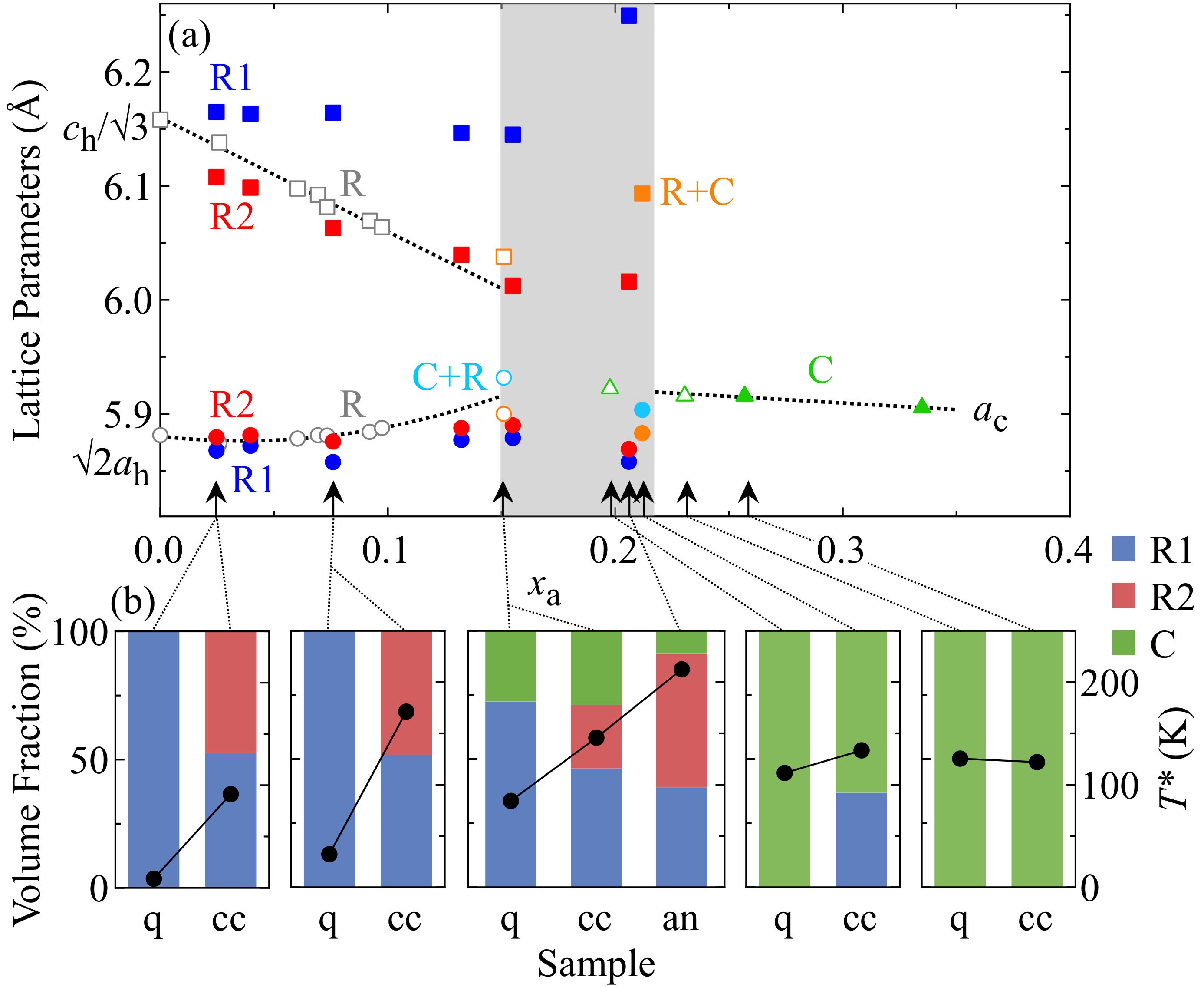}
\caption{(a) Lattice parameters of \GMT: The lattice parameters in the rhombohedral phase are given in their pseudo-cubic setting $\tilde{a}=\sqrt{2}a_{\rm h}$, $\tilde{c}=c_{\rm h}/\sqrt{3}$ for a better comparison. Circle symbols refer to $a_{\rm h}$, squares to $c_{\rm h}$, and triangles to the lattice parameter $a_{\rm c}$ in the cubic phase. Open symbols are used for quenched, filled symbols for controlled-cooled and annealed samples. In the high-\Tc\ phase of controlled-cooled samples, a multiphase scenario of two rhombohedral phase fractions is assumed (R1 and R2; blue and red). Upon increasing $\xa\gtrsim 0.080$, a cubic phase fraction starts to grow in addition; see text for details. The labels (C + R; cyan) and (R + C; orange) indicate two samples which could be refined best when assuming a cubic main phase plus a single rhombohedral phase fraction or a single rhombohedral main phase plus a cubic phase fraction, respectively. As in Fig.~\ref{fig1}, the gray-shaded area indicates the structural-transition range. Dotted lines are guides to the eyes. (b) Volume fraction of the different structural phases for selected quenched (``q''), controlled-cooled (``cc''), and annealed samples (``an''). The dotted arrows indicate their position, i.e., Mn concentrations \xa\ in panel (a). The black circle symbols are the respective ferromagnetic \Tc\ values (right axis), see text.}
\label{fig5}
\end{figure}

Except for the quenched sample with $\xa=0.151$, the lattice parameters for the quenched rhombohedral (low-\Tc) and the either quenched or controlled-cooled cubic specimen were already reported in our preceding paper Ref.~[\onlinecite{kriener16a}] (Fig.~3 therein). The SXRD patterns of these samples can be explained by either assuming rhombohedral or cubic phases. The corresponding lattice parameters are plotted in gray and green symbols in Fig.~\ref{fig5}~(a) and are labeled with a gray R or green C. 

\subsubsection{Controlled-cooled rhombohedral samples}
The Rietveld refinement and analysis of controlled-cooled rhombohedral samples (high-\Tc) were much more complicated due to the broadening and evolution of several peaks in the SXRD data with \xa. The aforementioned working hypothesis about the development of Mn-rich regions and hence a spatial variation of the rhombohedral distortion angle throughout the sample suggests a multi-phase approach to fit the experimental data. Motivated by the splitting of the rhombohedral $104_{\rm h}$ reflection into two broad peaks, we assume a simplified model for the analysis that there are two rhombohedral distortions in the samples characterized by two different but fixed distortion angles. As in our preceding paper, they are labeled R1 and R2. Hence we obtained two sets of $\tilde{a}_c$ and $\tilde{c}_c$ lattice parameters for each controlled-cooled sample below $\xa \approx 0.21$.\cite{R1R2comment} Those are plotted in blue (R1) and red (R2) in Figs.~\ref{fig5}~(a) and (b). The structural component R1 exhibits a stronger distortion, i.e., smaller Mn concentration and therefore corresponds to the aforementioned matrix of low-doped or even pristine GeTe, while the R2 component is less distorted, has a larger Mn concentration, and hence is closer to cubic structure. This component corresponds to the Mn-rich regions embedded in the surrounding matrix. 

Here it is important to emphasize that the observed splitting of the rhombohedral $104_{\rm h}$ reflection into two peaks R1 and R2 is \textit{not} due to the existence of an impurity phase nor a simple phase separation into two phases with sufficiently large size and well-defined Mn concentrations (a higher Mn-doped structural phase 1 and a lesser-doped phase 2). In the case of a conventional phase separation the reflections are expected to clearly split into sharp peaks which is apparently not the case for, e.g., the $104_{\rm h}$ reflection. Nevertheless we address our approach as two-phase model R1 + R2 for simplicity. 

We also note that the assumption of only two rhombohedral components, i.e., allowing for only two distinct distortion angles, reproduces qualitatively the two-peak structure of the rhombohedral $104_{\rm h}$ reflection, but does not always yield very good quantitative fits to the data, see Fig.~S5 in Ref.~\onlinecite{Suppl}, supporting the conclusion that these features are not due to a phase separation into sufficiently large domains. In reality, the rhombohedral distortion will rather change in a more continuous fashion according to the local Mn concentration. Moreover upon increasing \xa, the locally large concentration of Mn ions also leads to the formation of cubic phase fractions, further complicating the modeling. In our refinements, we had to expand the model from R1 + R2 to R1 + R2 + C to fit the data, starting from an annealed sample with $\xa = 0.106$ (not shown here, see Fig.~S7 in Ref.~[\onlinecite{Suppl}]). It is likely that these cubic phase fractions start to appear at doping concentrations for which the as-grown high-\Tc\ phase boundary in Fig.~\ref{fig1} starts to decrease, i.e., for $\xa \gtrsim 0.080$. For simplicity, the corresponding cubic lattice parameters are not plotted in Fig.~\ref{fig5}~(a).

\subsection{Volume fraction}
In Fig.~\ref{fig5}~(b), the volume fractions of 11 samples are shown. They are grouped into five histograms. Each histogram shows the ratio of the different phase fractions (left axis scale) for a quenched (labeled ``q'') and a controlled-cooled sample (``cc'') with similar Mn concentrations \xa. The central histogram contains data for three samples. Here the result for the annealed sample (``an'') which exhibits the highest \Tc\ is added. The black data points plotted on each bar indicate \Tc\ of that sample (right axis scale). The order of the histograms from left to right ``q'', ``cc'', and for the central one also ``an'', corresponds to the increased time which a sample was kept at elevated temperatures. In this sense the horizontal axis of each histogram represents the time for which the spinodal decomposition in a sample could proceed. The arrows connecting Figs.~\ref{fig5}~(a) and (b) indicate the respective Mn concentrations of the samples summarized in panel (b). We will discuss the evolution of the different phase fractions from left to right with increasing \xa. In the Supplemental Material, a similar histogram plot including data for additional samples can be found, see Fig.~S7 in Ref.~[\onlinecite{Suppl}].

The first two histograms contain data of samples with Mn concentrations smaller than 0.080. For none of them, the assumption of a cubic phase fraction was necessary in our Rietveld refinements. Both quenched samples, which are characterized by a more homogeneous Mn distribution, consist of a uniformly distorted rhombohedral structure R, which we identify with R1 (blue). The two controlled-cooled samples are subject to the spinodal decomposition which leads to the formation of two differently-distorted phase fractions. Each of the two phase fractions R1 (blue) and R2 (red) accounts for roughly half of the sample volume, in qualitative  agreement with the observation that the rhombohedral $104_{\rm h}$ reflection has split into two (broad) peaks with comparable intensity in these samples, cf.\ Figs.~\ref{fig4}~(c) and (d). At the same time, \Tc\ (black circle symbols in the histograms) is strongly enhanced compared to the case of the quenched samples. This clearly indicates that phase fraction R2 is responsible for the emergence of high-\Tc\ values.

The third histogram summarizes the volume fractions of three samples, all of them located in the structural-transition range $0.15 \lesssim \xa \lesssim 0.22$. In the quenched sample (left bar), about 25\% of the sample volume have already changed into the cubic phase. The fact that the sample has not completely switched yet, i.e., that there is no sharp structural phase transition, implies that quenched samples also exhibit a slight Mn inhomogeneity. In those parts of the sample where the local Mn concentration is sufficiently large, the structure locally turns into the cubic phase, while the rest of the sample remains distorted. However, \Tc\ remains small in agreement with the absence of the less-distorted rhombohedral phase fraction R2. The controlled-cooled sample in the third histogram of Fig.~\ref{fig5}~(b) (central bar), which has a very similar \xa\ as the quenched sample, also exhibits a cubic volume fraction of about 25\%. As in the case of the controlled-cooled samples shown in the first two histograms, roughly half of the sample consists of the strongly-distorted and Mn-poor phase fraction R1. Naturally, the cubic phase starts to grow in the Mn-rich R2 regions of a sample and reduces their volume. At the same time, \Tc\ has decreased compared to the controlled-cooled sample in the second histogram, which underlines the direct correlation between the existence of R2 phase fractions and high-\Tc\ values. The third sample in this histogram has a somewhat larger \xa, exhibits the highest \Tc\ observed in this study, and was annealed for three weeks at 500~K. Apparently, this heat treatment reduced the cubic phase fraction to less than 10\% while R2 has grown (or regrown)\cite{Probcomment} and accounts again for approximately 50\% of the sample volume. As a result, \Tc\ is strongly enhanced. This can be directly seen in Fig.~\ref{fig4}~(h), where the overall shape became ``more'' rhombohedral again in spite of the large average Mn concentration. 

To check whether it is possible to extract more quantitative information about the Mn concentration in the different phase fractions, we tried  two additional Rietveld refinements on the ``maximum-\Tc'' sample: (i) The first approach is under the assumption that the phase fractions R1, R2, and C have the same Mn concentration \xa. (ii) In the second scenario it is implied that R2 adsorbs all Mn from R1, i.e., R1 consists only of pure GeTe. Both models yield similarly good descriptions of the SXRD data, see Fig.~S6 in Ref.~[\onlinecite{Suppl}]. The estimated structural parameters are almost identical. Hence it is not possible to distinguish these two approaches quantitatively and determine the Mn concentrations of the different phase fractions independently by means of Rietveld refinements. The main problem here is that the features from different phases are too contiguous and hence it is difficult to separate their intensities from each other.

The fourth histogram in Fig.~\ref{fig5}~(b) shows the situation for a quenched and a controlled-cooled sample located in the upper half of the gray-shaded structural-transition range. The quenched compound has already completely switched into the cubic phase. By contrast, the controlled-cooled sample exhibits still a sizable rhombohedral volume fraction R1 of more than 1/3. However, there is no indication of a differently-distorted second phase fraction R2 any more in agreement with the finding that \Tc\ has dropped strongly to a similar value as it is observed for the quenched sample.

The fifth histogram consists of data for a quenched and a controlled-cooled sample in the cubic part of the phase diagram $\xa \gtrsim 0.22$. There is no difference any more in the ferromagnetic phase between either heat-treated sample, and their \Tc\ values are very similar.

The finding that the occurrence of high-\Tc\ values is directly correlated to the existence of rhombohedral R2 phase fractions leads to the very interesting question about the role of the polar-distortion-induced Rashba-spin splitting in enhancing \Tc\ values in \GMT. It was reported recently that the giant Rashba splitting of the bulk bands in pristine GeTe survives against the Mn doping.\cite{krempasky16a} Since the ferromagnetism in \GMT\ is charge-carrier mediated, it is reasonable to expect that there is an effect on the ferromagnetic exchange due to the band structure. Apparently \GMT\ exhibits high-\Tc\ values as long as the polar structure is preserved in parts of a sample. This suggests that the Rashba-split band structure in this system is at least partially responsible for the occurrence or in favor of high-\Tc\ values. Further experiments as well as theoretical input are required and desirable to shed light on this intriguing issue.

\subsection{Annealing effect}\label{anneff}
As described in the last part of Sec.~\ref{optimized}, not only \Tc\ but also \MTK\ is affected in our heat-treatment experiments. There are several conceivable possibilities, which may change the magnetic moment at 7~T and 2~K upon annealing: (i) degradation of the sample by time / heat cycles, (ii) formation (or extinction) of interstitial Mn$^{2+}$ defects, which are known to play an important role in the case of the textbook-diluted-magnetic semiconductor (Ga,Mn)As,\cite{ksato05a,dietl15a} or (iii) emergence (or extinction) of direct antiferromagnetic Mn$^{2+}-$Mn$^{2+}$ exchange interaction. 

As for (i), we sometimes observe a change in the surface color of a sample. Especially after subsequent annealing steps, it turns blackish. But this was found only to affect the surface of a sample, and scratching or polishing yields again very shiny surfaces. Hence the bulk part of the sample does not degrade. It also happens that small pieces break off, or that a sample breaks into two or more pieces. Those pieces can show different \Tc\ and \MTK\ values, reflecting the inhomogeneous situation of annealed samples. But for single pieces, the changes of the magnetic moment at 7~T and 2~K are not systematically correlated with the course of the annealing experiments, i.e., \MTK\ is observed to decrease \textit{and} increase and hence the magnetic moment can recover. Therefore degradation can be excluded to be responsible for the changes in \MTK.

The second possibility, formation or extinction of interstitials, changes the number of Mn$^{2+}$ ions which form Mn$^{2+}$\,--\,Mn$^{2+}$ pairs with  possibly strong antiferromagnetic coupling, and hence reduce or increase the measured moment \MTK. We cannot fully exclude that there are some interstitials, but we speculate that they are of small significance since we only observe such changes in \MTK\ for larger \xa. If interstitials would play a major role, changes for all \xa\ throughout the phase diagram are expected.

On the other hand, scenario (iii) seems feasible: In the rhombohedral high-\Tc\ phase, the spinodal decomposition in \GMT\ forms Mn-rich regions. If the total number of Mn$^{2+}$ ions is still small, the number of near-neighbor Mn$^{2+}-$Mn$^{2+}$ pairs with antiferromagnetic coupling is also small. Hence, additional annealing of low-doped samples cannot enhance the Mn$^{2+}-$Mn$^{2+}$ pair formation due to the lack of sufficient Mn$^{2+}$ ions and \MTK\ remains unchanged. However, upon increasing \xa, more and more Mn$^{2+}$ ions will get close to each other and hence the number of antiferromagnetically coupled Mn$^{2+}-$Mn$^{2+}$ pairs increases. This suggests that upon increasing \xa, the moment \MTK\ underestimates the real, i.e., chemical Mn concentration more and more due to increasing antiferromagnetic contributions. When annealing a sample, the Mn distribution changes due to the spinodal decomposition and therefore the antiferromagnetic contribution may vary in an unpredictable way. Thus, the antiferromagnetic contribution in a certain sample will vary, and the larger \xa\ becomes, the larger might be the change in \MTK\ before and after certain heat-treatment procedures. We note that there is no direct correlation between the size of the weakly-distorted phase fraction R2 and the saturation moment \MTK, cf.\ Figs.~\ref{fig5}~(b) and S7 in Ref.~\onlinecite{Suppl}.

 The remaining open question is why the experimentally observed changes in \MTK\ are largest in the cubic phase where there is no apparent spinodal decomposition at work and which is expected to exhibit a rather homogeneous Mn distribution. Here we can only speculate that there is also a slight Mn inhomogeneity present even in the supposedly homogeneous cubic phase. This could cause changes in \MTK\ because the overall Mn doping level is already so large that even a small Mn redistribution modifies the near-neighbor antiferromagnetic couplings. Since the phase fraction R2 does not form any more, \Tc\ is not much affected. When comparing the higher-angle SXRD results for a quenched ($\xa = 0.231$) and a controlled-cooled cubic sample ($\xa = 0.257$), it was found that the SXRD peaks in the data of the quenched sample are slightly sharper than those measured on the controlled-cooled sample, cf.\ Fig.~S9 in the Supplemental Material Ref.~[\onlinecite{Suppl}]. This could be a subtle indication of a slightly more inhomogeneous situation in controlled-cooled samples even in the cubic phase.

Finally, we comment on the interplay of the different phase fractions R1, R2, and C. In addition to the spinodal decomposition as driving force, there must be an interaction at work between the R1 and R2 phase fractions which allows the Mn-rich R2 fraction to remain rhombohedral even when the R2 phase is getting very close to cubic symmetry. This might be due to strong strain caused by the surrounding strongly-distorted Mn-poor R1 fraction. In other words, the accumulation of Mn in the R2 regions requires a stronger distortion in the surrounding R1 matrix. It is noted that the XRD data in Figs.~\ref{fig2} and \ref{fig4} suggest that R1 is at least partly more strongly distorted than pristine GeTe. In this sense, the R2 regions draw the Mn ions out of the R1 matrix and a subtle balance forms between the two phase fractions as the spinodal decomposition proceeds. The more Mn is incorporated into the R2 regions, the higher is \Tc. As soon as cubic phase fractions appear, R2 looses weight and hence they are responsible for the suppression of \Tc\ in controlled-cooled samples which leads to the dome-shape of \Tc. 
To achieve high values of \Tc, large Mn concentrations and hence larger mutual strain in the interplay of R1 and R2 are required. In this picture, the ideal high-\Tc\ phase line is the strongly increasing initial dotted line in the phase diagram in Fig.~\ref{fig1} up to the maximum of the dome at around $\xa \approx 0.080$, and then it keeps increasing linearly into the structural-transition range along the dotted line labeled ``annealed''. One may speculate that without structural phase transition, even higher-\Tc\ values would be feasible. The experimental finding that \Tc\ values above 200~K are not necessarily reproducible suggests that a certain initial Mn distribution is at least helpful or necessary in achieving such high-\Tc\ values. Hence, it is plausible to assume that the initial Mn distribution partly determines the detailed interplay between R1, R2, and C in a sample. To further investigate this issue, it would be very interesting to measure the Mn distribution directly by a local probe with a spatial resolution of $\sim 10$~nm.

\section{Summary}
\label{summary}
In conclusion we presented a comprehensive study of the structural evolution in \GMT\ with $x$, complementing our earlier work, where we reported that there are two different ferromagnetic phases in the rhombohedral part of the phase diagram. Which phase is realized depends on the heat treatment of the sample: controlled cooling (high \Tc) vs.\ quenching (low \Tc) from high temperatures. Here we could show that in the high-\Tc\ phase, differently rhombohedrally-distorted phase fractions develop with different Mn-doping levels. The less distorted Mn-rich fractions are responsible for the occurrence of high-\Tc\ values. Upon further doping, cubic phase fractions also emerge, and \Tc\ is suppressed again. The underlying mechanism is a spinodal decomposition and a complex interplay between the involved rhombohedral and cubic phase fractions. Moreover, we successfully demonstrated that it is possible to achieve even higher-\Tc\ values by optimizing the heat treatment. Here, a maximum bulk \Tc\ of about 214~K was found for $\xa=0.177$ after annealing the sample at 500~K for three weeks. This exceeds the transition temperatures in prototypical (Ga,Mn)As and many other diluted magnetic semiconductors. The comparison of  high-resolution synchrotron XRD data identified the increase of the less-distorted rhombohedral phase fractions accompanied by the reduction of the cubic phase fractions upon the annealing-induced enhancement of \Tc, clearly demonstrating the importance of the less-distorted rhombohedral phase fractions.
 

\begin{thebibliography}{65}
\bibitem[{xac()}]{xacomment}
\bibinfo{howpublished}{{For simplicity, the index ``a'' is omitted in the
  chemical formula of \GMT.}}
\bibitem[{UCe()}]{UCellcomment}
\bibinfo{howpublished}{{For a sketch of the rhombohedral and cubic unit cells
  of GeTe, see Fig.~1 in our preceding publication
  Ref.~\onlinecite{kriener16a}.}}
\bibitem[{XRD({\natexlab{a}})}]{XRDtimecomment1}
\bibinfo{howpublished}{{After stabilizing the temperature, the system was
  automatically realigned to compensate for the temperature-change-induced
  expansion of the components. This procedure took approximately
  10\,--\,15~minutes. Then the first XRD pattern was taken and subsequently we
  waited for 30~minutes. The alignment sequence was carried out prior to each
  XRD scan even when the temperature remained unchanged. Each XRD scan itself
  took about 10~minutes for a $2\theta$ range of $20^{\circ}-120^{\circ}$.}}
\bibitem[{XRD({\natexlab{b}})}]{XRDtimecomment2}
\bibinfo{howpublished}{{Between the first, second, and third scan, the waiting
  time was 30~minutes (plus alignment). After the third scan, the waiting time
  was increased to 60~minutes (plus alignment).}}
\bibitem[{Tan({\natexlab{a}})}]{Tanealcomment1}
\bibinfo{howpublished}{{Each time the samples were heated up into the cubic
  phase before annealing them at 500~K to start all annealing experiments from
  a homogeneous Mn distribution.}}
\bibitem[{Tan({\natexlab{b}})}]{Tanealcomment2}
\bibinfo{howpublished}{{The optimum annealing temperature $T_{\rm an}^{\rm
  opt}=500$~K was determined by using a sample with $\xa = 0.069$. Since
  annealing at this temperature enhanced \Tc\ above 200~K for $\xa = 0.206$,
  there seems to be no strong dependence of $T_{\rm an}^{\rm opt}$ on \xa.}}
\bibitem[{R1R()}]{R1R2comment}
\bibinfo{howpublished}{{In Fig.~3 of Ref.~\onlinecite{kriener16a}, we also
  presented lattice constants for the controlled-cooled sample with
  $\xa=0.076$. In the present paper, the SXRD data of this sample is shown in
  Fig.~\ref{fig4}~(d). The lattice constants shown in
  Ref.~\onlinecite{kriener16a} were estimated from the peak positions. The
  lattice constants for this sample given in this paper are deduced from a
  Rietveld refinement and are somewhat smaller than the crude estimate made in
  our previous publication.}}
\bibitem[{Pro()}]{Probcomment}
\bibinfo{howpublished}{{The data shown in the histograms were measured on
  different samples. For the SXRD measurements, the samples have to be pulverized after annealing. Therefore, we could not measure the respective SXRD profiles on the same sample after each annealing step.}}
\end{thebibliography}


\section*{Acknowledgments}
This work is partly supported by a Grants-in-Aid for Scientific Research (S) from the Japan Society for the Promotion of Science (JSPS, No. 24224009). MK is supported by a Grants-in-Aid for Young Scientists (B) (JSPS, KAKENHI No. 25800197) and by a Grants-in-Aid for Scientific Research (C) (JSPS, KAKENHI No. 15K05140). The synchrotron radiation experiments were performed at BL44B2 in SPring-8 with the approval of RIKEN (Proposal No.\ 20150045 and 20160006).

\end{document}